\documentclass[preprint,amsmath,nofootinbib,twoside]{revtex4}
\usepackage{feynmp}
\usepackage{graphicx}
\newcommand{\twi}{\widetilde}
\newcommand{\smass}[1]{m^2_{\twi #1}}
\DeclareMathOperator{\casim}{C_2^n}

\begin{document}

\title{Hidden sector renormalization of MSSM scalar masses}
\date{\today}

\author{Andrew G. Cohen}
\email{cohen@bu.edu}
\author{Tuhin S. Roy}
\email{tuhin@bu.edu}
\author{Martin Schmaltz}
\email{schmaltz@bu.edu}
\affiliation{Physics Department,  Boston University\\
  Boston, MA 02215, USA}
\preprint{BUHEP-06-08}

\begin{abstract}
  Running of gauge couplings in the MSSM from a unified value at high
  energies leads to a successful prediction of the weak mixing angle.
  Supersymmetric models at the TeV scale may contain further hints of
  high scale physics, such as the pattern of superpartner masses when
  evolved from the TeV scale to a high scale using the renormalization
  group.  This running is traditionally assumed to be independent of
  effects in the hidden sector. In this paper we re-examine this
  assumption, and conclude that the predictions for scalar masses may
  depend sensitively on the details of the mechanism of supersymmetry
  breaking. We identify mass relations that persist even when such
  effects are taken into account.
\end{abstract}

\maketitle

\section{Introduction}
\label{sec:introduction}

Weak scale supersymmetry is an attractive scenario for physics at the
TeV scale. The minimal supersymmetric standard model (MSSM) with
R-parity stabilizes the Higgs mass against radiative corrections,
predicts weakly interacting dark matter, and satisfies constraints
from precision electroweak measurements which rule out many other
extensions of the standard model.  With the additional
assumptions of dynamical supersymmetry breaking in a hidden sector and
flavor universal mediation of this breaking, the model also explains
the enormous hierarchy between the electroweak scale and the Planck
scale by dimensional transmutation while accommodating the constraints
from flavor changing neutral currents (FCNCs).

But an especially compelling feature of the MSSM is its correct
prediction of the weak mixing angle from gauge coupling unification
\cite{Dimopoulos:1981zb,Georgi:1974yf,Dimopoulos:1981yj}.  This
prediction supports the notion of a ``desert'' between the weak scale
and the unification scale at $M_{\text{GUT}} \sim 2 \times 10^{16}$
GeV, and raises the hope that TeV scale measurements may allow
indirect access to high scale physics through extrapolation with the
renormalization group.

Can we reliably predict other TeV scale observables from high scale
models? Or, working from the bottom up, can we probe high scale
physics through TeV scale measurements? Previous work has indicated
that superpartner masses may be just such observables
\cite{Inoue:1982pi,Ellis:1982fc,Alvarez-Gaume:1983gj,
Martin:1997ns,Ramond:1999vh,Weinberg:2000cr,Baer:2004rs,Terning:2006bq,
Drees:2004jm,Aguilar-Saavedra:2005pw}.%
\footnote{See \cite{Tsukamoto:1993gt} for a study of whether superpartner
mass relations can be tested with sufficient precision at colliders.}
These investigations imply that superpartner masses depend on the
means by which SUSY breaking is communicated to the visible world, but
not on the details of the hidden sector itself.  If correct, this
would suggest that models with simple messenger physics are highly
predictive. For this reason substantial effort has gone into refining
\cite{Martin:1993zk,Jack:2003sx} and even automating
\cite{Paige:2003mg,Allanach:2001kg,Djouadi:2002ze,Porod:2003um} the
renormalization group evolution of superpartner masses.

In this paper we demonstrate that this conventional wisdom may be
wrong.  While MSSM evolution of gaugino masses is reliable, we find
that the renormalization group evolution of MSSM scalar masses may
depend strongly on the unknown dynamics in the hidden sector. The
essential point is simple. Even though the visible and hidden sectors
are only coupled through higher dimensional operators, and therefore
renormalizable couplings in the two sectors run independently, the
higher dimensional operators themselves are renormalized by both
sectors. Furthermore, the renormalization of these operators does not
factor into a visible and a hidden contribution, as would na\"\i{}vely
seem to be the case, because of an additive contribution to scalar
masses from couplings of MSSM gauginos to the hidden sector. Therefore
hidden sector interactions must be taken into account when computing
the renormalization of MSSM scalar masses. This may introduce
dependence on new (and unknown) parameters in predictions for scalar
masses. In some cases these effects are small, while in others they
can even dominate over the usual one-loop MSSM running.  In all cases,
this dependence can be summarized in terms of a few new parameters,
and model-independent predictions remain.

Before demonstrating this result in a simple toy model we summarize
the main consequences of hidden sector contributions to visible sector
running. For simplicity we focus on the renormalization of first and
second generation scalar masses in this paper wherein the MSSM Yukawa
couplings can be neglected.

We begin by considering grand unified models (and high-scale messenger
sectors) wherein the operators responsible for scalar masses have
unified coefficients at $M_{\text{GUT}}$.  Ignoring hidden sector
interactions, MSSM renormalization would then give simple mass
relations at the TeV scale which follow from unification at the GUT
scale. We find that hidden sector renormalization can greatly modify
these relations. However, one linear combination of scalar masses runs
independently of hidden sector effects at one loop order in visible
(MSSM) couplings and to all orders in hidden sector couplings. This
allows us to predict
\begin{equation}
  \label{eq:1}
   m_{\twi Q}^2-2 m_{\twi U}^2 + m_{\twi D}^2 -m_{\twi L}^2 + 
   m_{\twi E}^2 = 0   
\end{equation}
at the TeV scale. This result holds for any unified theory in which no
hypercharge D-term is generated at the GUT scale.

Hidden sector running can also be important in models of gauge
mediation \cite{Dine:1981za,Dimopoulos:1981au,Dine:1981gu,
Nappi:1982hm,Alvarez-Gaume:1981wy,Dimopoulos:1982gm,
Dine:1993yw,Dine:1994vc,Dine:1995ag,Giudice:1998bp}.
Again, if the hidden sector couplings are strong, scalar mass
predictions may be significantly modified. Nonetheless, working at one
loop order in MSSM couplings and to all orders in the hidden sector,
there are two linear combinations of first generation scalar masses
for which hidden sector effects drop out. In addition to \eqref{eq:1}
we predict
\begin{equation}
  \label{eq:2}
  3(m_{\twi D}^2 - m_{\twi U}^2) + m_{\twi E}^2 = 0
\end{equation}
at the TeV scale, in models of gauge mediation.

Hidden sector models in which strong coupling persists over a large
range of scales often have large hidden sector anomalous dimensions
\cite{Luty:2001jh,Dine:2004dv,Ibe:2005pj,Schmaltz:2006qs}.  The
running from the hidden sector may then drive scalar masses to values
which are hierarchically different from gaugino masses at a scale
somewhere between the GUT and TeV scales.  In the case where these
anomalous dimensions cause scalar masses to become very small at the
intermediate scale, they are subsequently regenerated through gaugino
masses at energies below the intermediate scale. The superpartner
spectrum that results in this case is that of gaugino mediation
\cite{Kaplan:1999ac,Chacko:1999mi,Cheng:2001an}.

In the case where scalar masses are driven hierarchically larger than gaugino
masses the model fails to stabilize the electroweak scale and requires
fine-tuning.  Therefore this running places new constraints on hidden
sector dynamics. Theories which predict
very heavy scalars \cite{Wells:2003tf,Arkani-Hamed:2004fb} may
nevertheless be interesting in the context of anthropics and the
landscape.

There are also hidden sector models with only very weak
interactions. In such models the effects discussed in this paper may
be negligible.  Examples include the Polonyi model, and supergravity
models in which supersymmetry is broken only by moduli with couplings
suppressed by $M_{\text{Pl}}$.

The renormalization effects we are describing may be understood in a
simple toy model. Consider a hidden sector with a single chiral
superfield $X$ with superpotential interaction%
\footnote{This example is too simple to spontaneously break
  supersymmetry. We use it only to demonstrate the origin of
  renormalization effects from the hidden sector. A similar analysis
  applies in complete O'Raifeartaigh models.}
\begin{equation}
  \label{eq:3}
  \mathcal{W}_{h} =\frac{\lambda}{3!} X^3 \ .  
\end{equation}
We couple $X$ to a single generation of MSSM matter fields
$\Phi_i=[Q,U,D,L,E]_i$ and $SU(3)\!\times\! SU(2)\!\times\! U(1)$
gauge fields $W_n$ with the usual non-renormalizable interactions
produced at the messenger scale $M$:
\begin{equation}
  \label{eq:4}
  \int\! d^4\theta\, k_i \frac{X^\dagger X}{M^2} \Phi_i^{\dagger}
  \Phi_i + \int\! d^2\theta\, w \frac{X}{M} W_n W_n 
\end{equation}
For definiteness, we consider a grand unified model with a messenger
scale $M$ at or above the GUT scale. Unification then predicts that
the complex coefficient $w$ at the messenger scale is independent of
the standard model gauge group.  A non-renormalization theorem for
$w$, which holds in holomorphic renormalization
schemes\footnote{Throughout this paper we adopt a holomorphic scheme for
  all hidden sector fields as well as the MSSM gauge fields
  $W_{n}$. However we use canonically normalized MSSM matter and Higgs
  fields to more easily identify scalar masses.}, implies that $w$
remains independent of gauge group at all scales.  Assuming that the
hidden sector field acquires a supersymmetry breaking $F$-component
$\langle X\rangle \vert_{F} = F$ at the intermediate scale
$M_{\text{int}}$, the gaugino masses are given by the universal factor
$w F/M$ times gauge couplings squared.

In the following we will not distinguish the $X$-mass scale from the
scale at which supersymmetry breaks, $M_{\text{int}} \sim
10^{11}\text{ GeV}$.  This assumption, easily relaxed, introduces no
essential changes in our analysis while substantially easing the
notation.

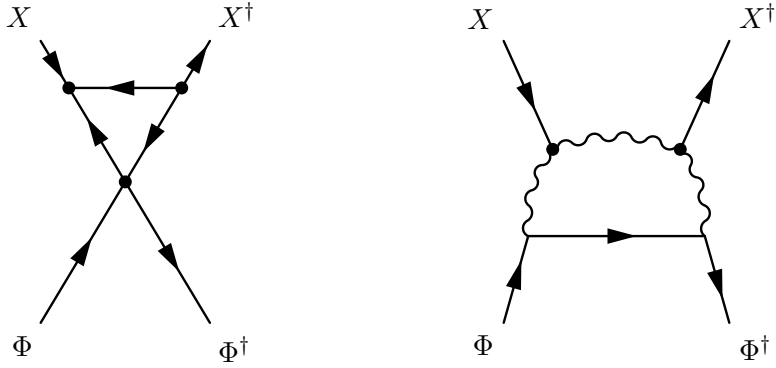
\begin{figure}[t]
\setlength{\unitlength}{1.5 mm}
\begin{fmffile}{figure1}
  {
    \begin{fmfgraph*}(15,25) 
      \fmfstraight
      \fmfleftn{i}{2}   \fmfrightn{o}{2}  
      \fmf{phantom,tension=1}{i1,v1,o1}
      \fmf{phantom}{i2,v1,o2}  \fmffreeze
      \fmf{fermion}{i1,v1,v2}
      \fmf{fermion}{v3,v1,o1}
      \fmf{fermion,tension=0}{v3,v2}
      \fmf{fermion,tension=2}{v3,o2}
      \fmf{fermion,tension=2}{i2,v2}
      \fmfdotn{v}{3}
      \fmflabel{$X$}{i2}
      \fmflabel{$X^\dag$}{o2}
      \fmflabel{$\Phi$}{i1}
      \fmflabel{$\Phi^\dag$}{o1}
    \end{fmfgraph*}
  }   \hfil
  {
    \begin{fmfgraph*}(20,25) 
      \fmfstraight
      \fmfleftn{i}{2}   \fmfrightn{o}{2} 
      \fmf{phantom}{i1,v1,v2,t1}
      \fmf{phantom}{t1,v3,v4,o1}  
      \fmf{fermion,tension=.8}{i2,v2}  
      \fmf{fermion,tension=.8}{v3,o2}       \fmffreeze
      \fmf{fermion}{i1,v1,v4,o1} 
      \fmf{photon,left=.2}{v1,v2,v3,v4}
      \fmfdot{v2,v3}
      \fmflabel{$X$}{i2}
      \fmflabel{$X^\dag$}{o2}
      \fmflabel{$\Phi$}{i1}
      \fmflabel{$\Phi^\dag$}{o1}
    \end{fmfgraph*}
  }
\end{fmffile}
\caption{Renormalization of the operators responsible for scalar masses.}
\label{fig:1}
\end{figure}

The renormalization of the coefficients $k_i$ which determine the soft
masses of the visible sector scalars $\phi_i$ is more interesting.  At
one loop, the $k_i$ are renormalized by visible sector gauge
interactions and the hidden sector Yukawa coupling $\lambda$ (see
figure \ref{fig:1}), and satisfy the renormalization group equations
\begin{equation}
  \label{eq:5}
  \begin{split}
    \frac{d}{dt} k_i &= \frac{2 \lambda^*\!\lambda}{16 \pi^2}k_i 
    - \frac{1}{16\pi^{2}} \sum_n 8\casim(R_i)\, g_n^6\ w^*\! w \\
    &\equiv \gamma k_i 
    -  \frac{1}{16\pi^{2}} \sum_n 8\casim(R_i)\, g_n^6 G  \ .
  \end{split}
\end{equation}
Here $t=\ln(\mu/M)$, $\lambda$ is the (running) Yukawa coupling of the
hidden sector, and $\casim(R_i)$ are group theory coefficients for the
matter fields in representation $R_i$ of the $n$-th MSSM gauge
group. The standard model gauge couplings $g_{n}$ run according to the
usual MSSM RGEs.  This differential equation is readily solved:
\begin{equation}
  \label{eq:6}
  k_i(t)= \exp\left(\!-\!\int^{0}_{t}\!\! dt' \,\gamma(t')\!\right) k_i(0)+
  \frac{1}{16\pi^2} \sum_n 8 \casim(R_i) \int_{t}^{0} \!ds\,
  g_n^6(s) \exp\left(\!-\!\int^{s}_{t} \!\! dt'\,
    \gamma(t')\!\right) G\ .
\end{equation}
Note that with our definitions we are interested in $t<0$, the region
below the messenger scale.  The exponential factor in the first term
effectively rescales all scalar masses by a common factor; this can be
absorbed into the messenger scale boundary value for coefficients of
the operators responsible for the scalar masses, $k_i(0)$.  Since only
a common factor is involved, this rescaling preserves any
relationships that might be present at the high scale. For example, in
the case of unified boundary conditions, as would arise in $SO(10)$,
this rescaling preserves such unified boundary conditions.  The second
term, an additive contribution which exists even without hidden sector
renormalization, splits the scalar masses. If the anomalous dimension
$\gamma$ vanishes, the second term involves only standard model gauge
couplings and the parameter $w$ that determines the gaugino masses. In
this case the TeV-scale scalar mass differences are related to the
gaugino masses.  However, when the coupling $\lambda$ is
non-negligible, the second term depends on hidden sector physics, and
spoils predictions that follow from these relations.

\begin{figure}[t]
  \begin{center}
    \includegraphics[width=.75\textwidth ]{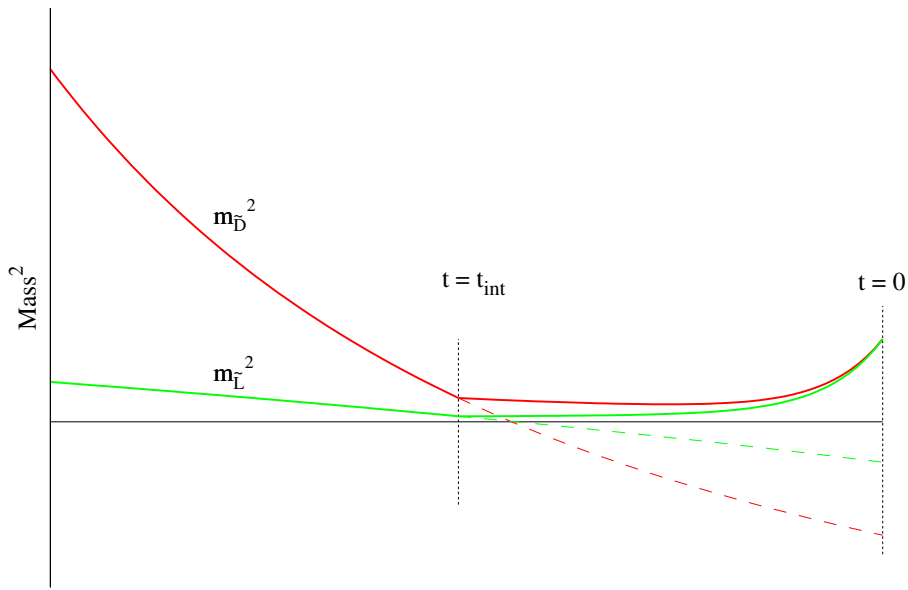}
  \end{center}
  \caption{Renormalization of $\tilde{D}$ and $\tilde{L}$ scalar
    masses squared. The solid curves show the renormalization
    including hidden sector effects. The dashed curves show these
    masses with the same values at 1 TeV run to higher scales without
    hidden sector effects. Note that ignoring hidden sector effects in
    the running would not show unification at $t=0$.}
  \label{fig:2}
\end{figure}

Note also that the hidden sector contributions to the second term
cannot be absorbed into a change of \textit{unified} boundary
conditions at the GUT scale. The example plotted in Figure
\ref{fig:2}, where we have chosen a very large (positive) value for
the hidden sector anomalous dimension $\gamma$, provides a good
illustration.  In this case, the hidden sector renormalization
strongly suppresses all scalar masses at the intermediate
scale. Clearly, this cannot result from MSSM running with any unified
boundary condition for the $k_i(0)$ at the GUT scale.

In general, the best we can do is parameterize the masses at the
intermediate scale $t_{\text{int}}=\ln (M_{\text{int}}/M)$ in terms of
four moments
\begin{align}
  \label{eq:7}
  N_0 &= \frac{\lvert F\rvert^{2}}{M^{2}}
  \exp\left(-\int_{t_{\text{int}}}^{0}\!  dt'\,
    \gamma(t')\right) k(0) \\
  N_n &= \frac{\lvert F\rvert^{2}}{M^{2}}\frac{1}{16\pi^2}
  \int_{t_{\text{int}}}^{0}\! ds\, \exp\left(-\int^{s}_{t_{\text{int}}}
    \!dt'\, \gamma(t')\right) g_n^6(s) G \qquad n=1,2,3
\end{align}
giving scalar masses
\begin{equation}
  \label{eq:8}
  m^{2}_{i}(t_{\text{int}}) =  k_i(t_{\text{int}})\frac{\lvert F\rvert^{2}}{M^{2}}
  =N_0 + \sum_{n=1}^3  8 \casim(R_i)\ N_{n} 
  \ .
\end{equation}
We have assumed a unified boundary condition at the messenger scale,
$k_{i}(0) = k(0)$.  At one loop the MSSM RGE coefficients $8\,
\casim(R_i)$ are
\begin{equation}
  \label{eq:9}
  \begin{array}{l|ccr}
    & SU(3) & SU(2) & U(1)\\ \hline
    {Q^{\phantom |}_{\phantom |} \ } & {32}/{3} & 6 & {2}/{15}\\
    {U{\phantom |}_{\phantom |}} & {32}/{3} & 0 & {32}/{15} \\
    {D{\phantom |}_{\phantom |}} & {32}/{3}& 0 & {8}/{15} \\
    {L{\phantom |}_{\phantom |}} & 0 & 6 & {6}/{5} \\
    {E{\phantom |}_{\phantom |}} & 0 & 0 & {24}/{5} 
  \end{array}
\end{equation}

To obtain predictions for the physical scalar masses at 1 TeV, we
should further evolve these intermediate scale masses using the MSSM
RGEs. At one-loop this evolution has the same form as \eqref{eq:5}
(without the hidden sector effects) and may therefore be absorbed into
the moments $N_{n}$.  Inclusion of higher order MSSM running would
require evolving these intermediate scale values down to the TeV scale
using the higher-loop MSSM RGEs.

It is easy to check that the combination of masses in \eqref{eq:1}
vanishes if the hypercharge D-term vanishes at the GUT scale, as is
required by the non-Abelian gauge invariance of the GUT group
containing hypercharge. If we were to allow a non-vanishing
hypercharge D-term, as may be the case in GUT models with unified
gauge groups of higher rank
\cite{Drees:1986vd,Kolda:1995iw,Baer:1999mc}, then the prediction
\eqref{eq:1} would be lost.

How significant are the contributions from the hidden sector running?
For strongly coupled theories modifications of $N_0$ may be of order
one or larger, whereas for weakly coupled theories they are suppressed
by a loop factor times a log compared to the tree level
value. However, contributions to $N_0$ can be absorbed into the
unknown UV boundary condition $k_i(0)$, and are therefore of less
interest%
\footnote{In the context of the MSUGRA model the hidden sector
  renormalization of $N_0$ is equivalent to a rescaling of $m_0$
  relative to $m_{1/2}$ and $A$-terms at the GUT scale. Wave function
  renormalization of hidden sector fields yields a similar universal
  rescaling of all MSSM soft masses relative to the gravitino mass.
  This point has been emphasized in \cite{Dine:2004dv}.}.

Note that even though the $N_n$ are formally suppressed by a loop
factor relative to $N_0$, their contributions to scalar masses in the
MSSM are numerically of similar size and split the scalar masses
significantly. Unlike contributions to $N_{0}$, hidden sector
modifications of the $N_n$ cannot in general be absorbed into UV
boundary conditions without destroying relations imposed by the high
scale physics. For hidden sectors which become strongly coupled at the
intermediate scale, as might be expected in dynamical supersymmetry
breaking models, the effects on the $N_n$ are $\mathcal{O}(1)$.
Hidden sectors which remain strongly coupled for a range of scales can
have even larger effects. But even weakly coupled hidden sectors may
have noticeable impact.  For example, if the hidden sector couplings
are as weak as the MSSM couplings, then the leading non-universal
contributions to scalar masses are of order $N_n^{\text{MSSM}}$ times
a (hidden sector) loop factor with a logarithmic enhancement from the
running; this is larger by a logarithm relative to the usual two-loop
MSSM contributions.

Measuring the second generation scalar masses is unlikely to give us
independent information on high scale physics.  This is because limits
on flavor changing neutral currents already require the first and
second generation scalars to be nearly degenerate.  This degeneracy,
once imposed at the high scale, is preserved by hidden sector
renormalization, and therefore the second generation scalar masses are
expected (by flavor universality) to be the same as first generation
masses. The third generation is more interesting and is discussed in
section \ref{sec:third-generation}.

\section{General Hidden Sectors}
\label{sec:gener-hidd-sect}

The above argument is straightforward to generalize to hidden sectors
with multiple chiral and vector superfields and more complex
interactions. We continue to work to one-loop order in visible sector
couplings but incorporate hidden sector couplings to all orders in
perturbation theory. To begin, let us introduce an efficient notation
for hidden sector operators. We label all gauge invariant real
superfield operators by $V_v$ and all chiral operators by $X_x$.  In
principle $v$ and $x$ enumerate a very large number of operators but
in practice only a few of them have small enough dimension to be
relevant. For convenience we will assume that the $X_{x}$ are
normalized (with powers of the messenger scale) to have engineering
dimension one, while the $V_{v}$ have engineering dimension two.  In
the example of the previous section we have $X_1=X$ and $V_1=X^\dagger
X$. Clearly, there is an arbitrariness in the choice of basis for
these operators, and $V_v$ for different $v$ may mix under
renormalization. Renormalization due to hidden sector interactions
needs to be taken into account from the messenger scale down to an
intermediate scale $t_{\text{int}}$ at which the hidden sector
dynamically breaks SUSY\footnote{We have again ignored any difference
  between the scale at which SUSY breaks and the masses of the hidden
  sector fields. Relaxing this assumption is straightforward.}. At
$t_{\text{int}}$ we replace the auxiliary components of the hidden
sector operators $V_v$ and $X_x$ by their expectation values
\begin{equation}
  \label{eq:10}
  \langle V_v \rangle\vert_D = D_v \quad\quad\quad
  \langle X_x \rangle\vert_F = F_x \ .  
\end{equation}

General hidden-visible couplings (relaxing the assumption of unified
gaugino mass boundary conditions) take the form
\begin{equation}
  \label{eq:11}
  \int\!d^4\theta\, k_{vi}\, \frac{V_v}{M^{2}} \Phi_i^{\dagger} \Phi_i
  +  \int\!d^2\theta\, w_{xn} \frac{X_x}{M} W_n W_n 
\end{equation}
where we have suppressed all indices labeling MSSM generations.  In
general, the scalar mass operators may be different for the three
generations, in which case the coefficients $k$ carry additional
flavor indices. Continuing to suppress such indices, the superpartner
masses at the intermediate scale are given by
\begin{align}
  \label{eq:12}
  \text{gaugino:}& & &M_n=
  \left(
    \sum_{x}w_{xn} \frac{F_x}{M} 
  \right)g^2_n(t_{\text{int}}) \\
  \text{scalar:}& & &m_i^2= 
  \sum_{v}\frac{D_v}{M^{2}} k_{vi}(t_{\text{int}})\  .
\end{align}
\begin{figure}[t]
\setlength{\unitlength}{1.5 mm}
\begin{fmffile}{figure2}
  {
    \begin{fmfgraph*}(15,20) 
      \fmfstraight
      \fmfbottom{i1,o1} \fmftop{t1}
      \fmf{fermion}{i1,v1,o1}
      \fmf{dbl_wiggly,label=$v$,l.side=left,tension=2}{v2,t1} 
      \fmf{dbl_wiggly,label=$v'$,l.side=left,tension=2}{v1,v2}      
      \fmfdotn{v}{2}  \fmfblob{.3w}{v2}
    \end{fmfgraph*}
  }\hfil
  {
    \begin{fmfgraph*}(15,20)   
      \fmfstraight
      \fmfbottom{i1,o1}\fmftop{t1}
      \fmfpolyn{shaded,smooth,pull=1.5,tension=1.1}{H}{3}
      \fmf{phantom}{i1,v1,v2,H1}
      \fmf{phantom}{H2,v3,v4,o1} 
      \fmf{phantom,tension=2}{t1,H3}  \fmffreeze
      \fmf{fermion}{i1,v1,v4,o1}
      \fmf{photon,left=.2}{v1,v2,v3,v4}
      \fmf{double,label=$x$,l.side=left}{v2,H1}  
      \fmf{double,label=$x'$,l.side=left}{H2,v3} 
      \fmf{dbl_wiggly,label=$v$,tension=2}{H3,t1} 
      \fmf{double,label=$x$,l.side=left}{v2,H1}  
      \fmf{phantom_arrow}{v2,H1}   
      \fmf{double,label=$x'$,l.side=left}{H2,v3} 
      \fmf{phantom_arrow}{H2,v3}
      \fmfdotn{v}{4}
    \end{fmfgraph*}
  } 
\end{fmffile}
\caption{Renormalization of the operators responsible for scalar
  masses, including arbitrary hidden sector effects. Single solid
  lines represent the MSSM fields $Q$, double wavy lines represent the
  vector operators $V$, and double straight lines represent the
  chiral fields $X$.}
\label{fig:3}
\end{figure}
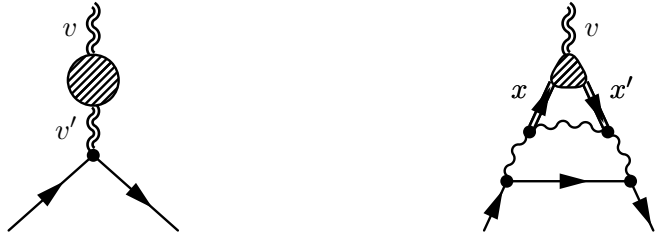
The couplings $w_{xn}$ are not renormalized in perturbation theory. The
$k_{vi}$ are renormalized by the diagrams in Figure \ref{fig:3}, where
arbitrary hidden sector renormalizations are included through the
blobs:
\begin{equation}
  \label{eq:13}
  \begin{split}
    \frac{d}{dt} k_{vi} &= \gamma_{vv'}\, k_{v'i} -
    \frac{1}{16\pi^2} \sum_n 8 \casim(R_i)\, g_n^6\, w^*_{xn}
    J_{vxx'}\, w_{x'n} \\
    &\equiv \gamma_{vv'}\, k_{v'i} - \frac{1}{16\pi^2} \sum_n 8
    \casim(R_i)\, g_n^6\, G_{vn} \ .
  \end{split}
\end{equation}
Repeated indices $v',x,x'$ are summed over, $\gamma_{vv'}(t)$ is the
anomalous dimension matrix of the operators $V_v$ in the absence of
visible sector interactions and $G_{vn}(t)\equiv w^*_{xn}
J_{vxx'}(t)w_{x'n}$ are three vectors of real functions (one vector
for each standard model gauge group) determined by hidden sector
interactions.  Note that the blob connecting vector operators $V$ with
chiral operators $X, X^{\dagger}$, represented by the $J$s, may
include disconnected components as well as connected ones.  In the
absence of hidden sector interactions the anomalous dimension matrix
vanishes, and $J$ simply relates the basis of free chiral
fields to the vector fields $X^{\dagger}X$ formed as products of these
chiral fields. For example the toy model of the previous section, with
only one chiral field $X$ and only one vector field $X^{\dagger}X$,
has a single blob, and $J=1$, $G_{1n}=w^*\!w$ at one-loop.

In the following, we will use a matrix notation for the indices $v,v'$
labeling the vector operators. Then the RGEs become identical in form
to the RGE of our simple toy model
\begin{equation}
  \label{eq:14}
  \frac{d}{dt} k_{i}=
  \gamma\, k_{i} - \frac{1}{16\pi^2} \sum_n 8\, \casim(R_i)\,
  g_n^6\  G_n 
\end{equation}
and it is straightforward to generalize the solution. The only new
feature is the appearance of path ordered exponentials to account for
any non-commutativity of the matrices $\gamma(t)$ at different $t$
\begin{multline}
  \label{eq:15}
  k_{i}(t)=  \text{P} \exp\left(-\int_{t}^{0}\!
    dt'\, \gamma(t')   \right)k_{i}(0) \\
  +\frac{1}{16\pi^2}\sum_n 8 \casim(R_i) \int_t^{0}\!
  ds \,  g_n^6(s) \text{P} \exp\left(-\int^{s}_t \!
    dt'\,\gamma(t') \right) G_n(s)  
\end{multline}
Multiplying by the (vector of) expectation values $D$ and assuming
unified boundary conditions for the $k_{i}(0)=k$ we obtain the scalar
masses at the intermediate scale
\begin{equation}
  \label{eq:16}
  m^2_i(t_{\text{int}})=N_0 +
  \sum_{n=1}^3 8\, \casim(R_i) N_{n} 
\end{equation}
in terms of the moments
\begin{align}
  \label{eq:17}
  N_0 &= \frac{D}{M^{2}}\
  \text{P}\exp\!\left(-\int_{t_{\text{int}}}^{0} \! dt'\,
    \gamma(t') \right) k \\
  N_n &= \frac{1}{16\pi^2}\frac{D}{M^{2}}\int_{t_{\text{int}}}^{0}\!
  ds\, \ g_n^6(s) \text{P}\exp\left(-\int^{s}_{t_{\text{int}}} \!dt'\,
    \gamma(t') \right) G_n(s), \quad n=1,2,3
\end{align}

\section{The Third Generation}
\label{sec:third-generation}

The analysis of the third generation scalar masses is similar,
although complicated by the presence of a large top Yukawa coupling
and possibly other interactions. The presence of this coupling
necessarily connects the renormalization group flow of the top and
bottom quarks with that of the Higgs scalars, and we therefore treat
the two Higgs doublets along with the five matter fields of the third
generation. This gives seven scalar masses in total.

The complete analysis, while straightforward, is messy, and we
postpone the details for a subsequent publication \cite{crs2}. However
the results are easily summarized. The most predictive case applies to
theories with a single, universal soft mass for all matter and Higgs
scalars, all Yukawa couplings small except for the top Yukawa, and
$A$-terms proportional to Yukawas (this is the case for minimal
supergravity). The common soft mass may be eliminated from predictions
by subtracting masses from the first generation. The presence of the
large top Yukawa, which introduces a new moment when hidden sector
renormalization is incorporated, requires forming linear combinations
that run independently of this coupling. This leads to six
predictions:

\begin{gather}
\label{eq:18}
  2 \smass{Q_{3}} -\smass{U_{3}}-\smass{D_{3}} -2 \smass{Q_{1}}
  +\smass{U_{1}}+\smass{D_{1}} = 0\\
\label{eq:19}
  2\smass{L_{3}} - \smass{E_{3}} -2\smass{L_{1}} + \smass{E_{1}} = 0\\
  \label{eq:20}
  2\smass{Q_{3}} - \smass{U_{3}}  - 2\smass{Q_{1}} + \smass{U_{1}}=
  0\\
\label{eq:21}
  \smass{E_{3}} -  \smass{E_{1}} = 0\\
\label{eq:22}
  3\smass{U_{3}}-3 \smass{D_{3}}+ 2\smass{L_{3}}-2 \smass{E_{3}}
  -2m^{2}_{H}+2 m^{2}_{\bar{H}} -3\smass{U_{1}}+3 \smass{D_{1}}-
  2\smass{L_{1}}+2 \smass{E_{1}}= 0\\
\label{eq:23}
  -3 \smass{D_{3}} -\smass{E_{3}} + 2 m^{2}_{\bar{H}} +3 \smass{D_{1}}
  -2\smass{L_{1}}+\smass{E_{1}} = 0
\end{gather}
Relaxing the condition of universal soft masses, or incorporating
large couplings leads to more parameters and hence fewer predictions.
For example if the Higgs soft masses differ from those of the matter
fields (but are the same for $H$ and $\bar{H}$) then the prediction
\eqref{eq:23} is lost. If the soft masses for $H$ and $\bar{H}$
differ, then \eqref{eq:22} is also lost. Independently, if the $\mu$
term arises from the Giudice-Masiero \cite{Giudice:1988yz} mechanism
($\mu H\bar H|_F \subset X^\dagger H\bar H|_D$) then both
\eqref{eq:22} and \eqref{eq:23} are lost.  The inclusion of singlet
neutrino fields with a large Yukawa coupling would eliminate
prediction \eqref{eq:19}.  Finally if $\tan\beta$ is large then the
predictions \eqref{eq:20} and \eqref{eq:21} are no longer valid.
Thus, \eqref{eq:18} remains as the only robust prediction.  The
others, \eqref{eq:19}-\eqref{eq:23} are still interesting as they
function as indirect probes.  For example, finding a violation of
\eqref{eq:19} would be indirect evidence for the presence of a singlet
neutrino with a large Yukawa coupling.

\section{Non-perturbative Renormalization}
\label{sec:non-pert-renorm}

Since we have already allowed for arbitrary functions $\gamma$ and
$G$, the renormalization group equation for the scalars does include
non-perturbative effects in the hidden sector. The gaugino mass
operators $X_x\, [\sum_n w_{xn} W_n W_n]$ may receive non-perturbative
renormalizations from the hidden sector. Any new terms which may be
generated from non-perturbative hidden sector dynamics can only depend
on the $W_n$ in the combinations $[\sum_n\! w_{xn} W_n W_n]$.
Furthermore, contributions which are relevant to gaugino masses
must be linear in $[\sum_n\! w_{xn} W_n W_n]$. It follows
that if gaugino masses were unified at some scale (\textit{i.e.\/} all
$w_x$ independent of $n$), then the equality of ratios
$M_1/g_1^2=M_2/g_2^2=M_3/g_3^2$
holds at all scales.

\section{Gauge Mediation}
\label{sec:gauge-mediation}

In models with gauge mediation supersymmetry breaking arises in the
hidden sector and is communicated to the visible sector via shared
gauge interactions. Usually the existence of messengers with standard
model quantum numbers and supersymmetric masses $M$ is
assumed. Integrating out these messengers generates the operators
coupling hidden sector fields to MSSM fields \eqref{eq:11} with
coefficients which are then calculable (for a given choice of
messengers). For our purposes we may be general and assume an
arbitrary messenger sector with MSSM fields coupling to the messengers
only through their gauge interactions. To leading (two-loop) order in
the MSSM interactions, and with arbitrary messenger and hidden sectors
this gives (see for example \cite{Weinberg:2000cr})
\begin{equation}
  \label{eq:36}
  k_{i}(0)=\sum_n 8\,\casim(R_i)\, K_{n}
\end{equation}
at the messenger scale $M$. Here we have suppressed indices $v$
labeling the hidden sector vector operators, and the functions $K_n$
parameterize the details of the messenger and hidden sectors.  Running
of the $k_{i}$ down to the intermediate scale $M_{\text{int}}$ is
governed by the same renormalization group equations as before,
\eqref{eq:13}.  The solution is
\begin{equation}
  \label{eq:37}
  m^2_i(t_{\text{int}})= \frac{D}{M^{2}}\
  \text{P}\exp\!\left(-\int_{t_{\text{int}}}^{0}  
    \! dt'\, \gamma(t') \right)  k_i(0) + \sum_{n=1}^3 8 \casim(R_i)
  N_{n}  
\end{equation}
The two terms may be combined: defining the new moments $\twi N_n$
\begin{equation}
  \label{eq:38}
  \twi N_{n}= \frac{D}{M^{2}}\
  \text{P}\exp\!\left(-\int_{t_{\text{int}}}^{0} \!dt'\, 
    \gamma(t') \right)  K_n +N_{n}
\end{equation}
we have
\begin{equation}
  \label{eq:39}
  m^2_i(t_{\text{int}})= \sum_{n=1}^3 8\, \casim(R_i)\, \twi N_{n}
\end{equation}
Thus the first generation scalar masses are given in terms of only
three unknown moments, leading to two predictions independent of the
hidden sector
\begin{gather}
  \label{eq:40}
 m_{\twi Q}^2-2 m_{\twi U}^2 + m_{\twi D}^2 -m_{\twi L}^2 + m_{\twi
   E}^2 =0 \\
3(m_{\twi D}^2 - m_{\twi U}^2) + m_{\twi E}^2  =0
\end{gather}
Since these combinations of masses are RG invariant at one-loop in
MSSM couplings (and to all orders in hidden sector interactions) these
predictions hold for the masses at all scales.  Inclusion of higher
order MSSM running would require evolving these intermediate scale
values down to the TeV scale with the two-loop MSSM RGEs. This
introduces a (weak) dependence on the unknown scale $M_{\text{int}}$
at the two-loop level.

\section{Conclusions}
\label{sec:conclusions}

We have demonstrated that interactions of the hidden sector introduce
uncertainties into the renormalization of MSSM scalar masses.  These
new effects \textit{cannot} be incorporated by simply rescaling the
coefficients of scalar mass operators without modifying UV coupling
relations.

Our results make testing unification in the pattern of scalar masses
at the TeV scale significantly more challenging.  Only when the hidden
sector is weakly coupled can the scalar masses be evolved to the high
scale without knowledge of hidden sector interactions. Without this
knowledge, the mass relation \eqref{eq:1} remains as the only
model-independent test of unification. In the case of gauge mediation
there is one further prediction, \eqref{eq:2}.

In a forthcoming paper we will present a detailed analysis of third
generation scalar masses in the presence of interacting hidden
sectors, and begin an exploration of specific classes of hidden
sectors which allow predictions in addition to \eqref{eq:1}.

\begin{acknowledgments}
  We thank Ami Katz for pointing out the possibility of probing
  dynamical $\mu$-term generation with hidden sector running.
  M.S. thanks Howie Baer for an inspiring conversation at Taco
  Bell. This work was supported in part by the Department of Energy
  under grant no. DE-FG02-01ER-40676, grant no. DE-FG02-91ER-40676 and
  an Alfred P. Sloan Research Fellowship.
\end{acknowledgments}

\bibliography{reference2}

\providecommand{\href}[2]{#2}\begingroup\raggedright\begin{thebibliography}{10}

\bibitem{Dimopoulos:1981zb}
S.~Dimopoulos and H.~Georgi, {\it Softly broken supersymmetry and su(5)},  {\em
  Nucl. Phys.} {\bf B193} (1981) 150.

\bibitem{Georgi:1974yf}
H.~Georgi, H.~R. Quinn, and S.~Weinberg, {\it Hierarchy of interactions in
  unified gauge theories},  {\em Phys. Rev. Lett.} {\bf 33} (1974) 451--454.

\bibitem{Dimopoulos:1981yj}
S.~Dimopoulos, S.~Raby, and F.~Wilczek, {\it Supersymmetry and the scale of
  unification},  {\em Phys. Rev.} {\bf D24} (1981) 1681--1683.

\bibitem{Inoue:1982pi}
K.~Inoue, A.~Kakuto, H.~Komatsu, and S.~Takeshita, {\it Aspects of grand
  unified models with softly broken supersymmetry},  {\em Prog. Theor. Phys.}
  {\bf 68} (1982) 927.

\bibitem{Ellis:1982fc}
J.~R. Ellis, L.~E. Ibanez, and G.~G. Ross, {\it Grand unification with large
  supersymmetry breaking},  {\em Phys. Lett.} {\bf B113} (1982) 283.

\bibitem{Alvarez-Gaume:1983gj}
L.~Alvarez-Gaume, J.~Polchinski, and M.~B. Wise, {\it Minimal low-energy
  supergravity},  {\em Nucl. Phys.} {\bf B221} (1983) 495.

\bibitem{Martin:1997ns}
S.~P. Martin, {\it A supersymmetry primer},
  \href{http://xxx.lanl.gov/abs/hep-ph/9709356}{ hep-ph/9709356}.

\bibitem{Ramond:1999vh}
  P.~Ramond,{\it Journeys Beyond The Standard Model}, .
  Reading, Mass., Perseus Books, 1999

\bibitem{Weinberg:2000cr}
S.~Weinberg, {\it The quantum theory of fields. vol. 3: Supersymmetry}, .
  Cambridge, UK: Univ. Pr. (2000) 419 p.


\bibitem{Baer:2004rs}
H.~Baer and X.~Tata, {\it Weak scale supersymmetry: From superfields to
  scattering events}, . Cambridge, UK: Univ. Pr. (2004) 537 p.

\bibitem{Terning:2006bq}
J.~Terning, {\it Modern supersymmetry: Dynamics and duality}, . Oxford, UK:
  Clarendon (2006) 324 p.

\bibitem{Drees:2004jm}
M.~Drees, R.~Godbole, and P.~Roy, {\it Theory and phenomenology of sparticles:
  An account of four- dimensional n=1 supersymmetry in high energy physics}, .
  Hackensack, USA: World Scientific (2004) 555 p.

\bibitem{Aguilar-Saavedra:2005pw}
J.~A. Aguilar-Saavedra {\em et~al.}, {\it Supersymmetry parameter analysis: Spa
  convention and project},  {\em Eur. Phys. J.} {\bf C46} (2006) 43--60,
  [\href{http://xxx.lanl.gov/abs/hep-ph/0511344}{ hep-ph/0511344}].

\bibitem{Tsukamoto:1993gt}
T.~Tsukamoto, K.~Fujii, H.~Murayama, M.~Yamaguchi, and Y.~Okada, {\it Precision
  study of supersymmetry at future linear e+ e- colliders},  {\em Phys. Rev.}
  {\bf D51} (1995) 3153--3171.

\bibitem{Martin:1993zk}
S.~P. Martin and M.~T. Vaughn, {\it Two loop renormalization group equations
  for soft supersymmetry breaking couplings},  {\em Phys. Rev.} {\bf D50}
  (1994) 2282, [\href{http://xxx.lanl.gov/abs/hep-ph/9311340}{
  hep-ph/9311340}].

\bibitem{Jack:2003sx}
I.~Jack, D.~R.~T. Jones, and A.~F. Kord, {\it Three loop soft running,
  benchmark points and semi- perturbative unification},  {\em Phys. Lett.} {\bf
  B579} (2004) 180--188, [\href{http://xxx.lanl.gov/abs/hep-ph/0308231}{
  hep-ph/0308231}].

\bibitem{Paige:2003mg}
F.~E. Paige, S.~D. Protopopescu, H.~Baer, and X.~Tata, {\it Isajet 7.69: A
  monte carlo event generator for p p, anti-p p, and e+ e- reactions},
  \href{http://xxx.lanl.gov/abs/hep-ph/0312045}{ hep-ph/0312045}.

\bibitem{Allanach:2001kg}
B.~C. Allanach, {\it Softsusy: A c++ program for calculating supersymmetric
  spectra},  {\em Comput. Phys. Commun.} {\bf 143} (2002) 305--331,
  [\href{http://xxx.lanl.gov/abs/hep-ph/0104145}{ hep-ph/0104145}].

\bibitem{Djouadi:2002ze}
A.~Djouadi, J.-L. Kneur, and G.~Moultaka, {\it Suspect: A fortran code for the
  supersymmetric and higgs particle spectrum in the mssm},
  \href{http://xxx.lanl.gov/abs/hep-ph/0211331}{ hep-ph/0211331}.

\bibitem{Porod:2003um}
W.~Porod, {\it Spheno, a program for calculating supersymmetric spectra, susy
  particle decays and susy particle production at e+ e- colliders},  {\em
  Comput. Phys. Commun.} {\bf 153} (2003) 275--315,
  [\href{http://xxx.lanl.gov/abs/hep-ph/0301101}{ hep-ph/0301101}].

\bibitem{Dine:1981za}
M.~Dine, W.~Fischler, and M.~Srednicki, {\it Supersymmetric technicolor},  {\em
  Nucl. Phys.} {\bf B189} (1981) 575--593.

\bibitem{Dimopoulos:1981au}
S.~Dimopoulos and S.~Raby, {\it Supercolor},  {\em Nucl. Phys.} {\bf B192}
  (1981) 353.

\bibitem{Dine:1981gu}
M.~Dine and W.~Fischler, {\it A phenomenological model of particle physics
  based on supersymmetry},  {\em Phys. Lett.} {\bf B110} (1982) 227.

\bibitem{Nappi:1982hm}
C.~R. Nappi and B.~A. Ovrut, {\it Supersymmetric extension of the su(3) x su(2)
  x u(1) model},  {\em Phys. Lett.} {\bf B113} (1982) 175.

\bibitem{Alvarez-Gaume:1981wy}
L.~Alvarez-Gaume, M.~Claudson, and M.~B. Wise, {\it Low-energy supersymmetry},
  {\em Nucl. Phys.} {\bf B207} (1982) 96.

\bibitem{Dimopoulos:1982gm}
S.~Dimopoulos and S.~Raby, {\it Geometric hierarchy},  {\em Nucl. Phys.} {\bf
  B219} (1983) 479.

\bibitem{Dine:1993yw}
M.~Dine and A.~E. Nelson, {\it Dynamical supersymmetry breaking at
  low-energies},  {\em Phys. Rev.} {\bf D48} (1993) 1277--1287,
  [\href{http://xxx.lanl.gov/abs/hep-ph/9303230}{ hep-ph/9303230}].

\bibitem{Dine:1994vc}
M.~Dine, A.~E. Nelson, and Y.~Shirman, {\it Low-energy dynamical supersymmetry
  breaking simplified},  {\em Phys. Rev.} {\bf D51} (1995) 1362--1370,
  [\href{http://xxx.lanl.gov/abs/hep-ph/9408384}{ hep-ph/9408384}].

\bibitem{Dine:1995ag}
M.~Dine, A.~E. Nelson, Y.~Nir, and Y.~Shirman, {\it New tools for low-energy
  dynamical supersymmetry breaking},  {\em Phys. Rev.} {\bf D53} (1996)
  2658--2669, [\href{http://xxx.lanl.gov/abs/hep-ph/9507378}{
  hep-ph/9507378}].

\bibitem{Giudice:1998bp}
G.~F. Giudice and R.~Rattazzi, {\it Theories with gauge-mediated supersymmetry
  breaking},  {\em Phys. Rept.} {\bf 322} (1999) 419--499,
  [\href{http://xxx.lanl.gov/abs/hep-ph/9801271}{ hep-ph/9801271}].

\bibitem{Luty:2001jh}
M.~A. Luty and R.~Sundrum, {\it Supersymmetry breaking and composite extra
  dimensions},  {\em Phys. Rev.} {\bf D65} (2002) 066004,
  [\href{http://xxx.lanl.gov/abs/hep-th/0105137}{ hep-th/0105137}].

\bibitem{Dine:2004dv}
M.~Dine {\em et~al.}, {\it Visible effects of the hidden sector},  {\em Phys.
  Rev.} {\bf D70} (2004) 045023,
  [\href{http://xxx.lanl.gov/abs/hep-ph/0405159}{ hep-ph/0405159}].

\bibitem{Ibe:2005pj}
M.~Ibe, K.~I. Izawa, Y.~Nakayama, Y.~Shinbara, and T.~Yanagida, {\it
  Conformally sequestered susy breaking in vector-like gauge theories},  {\em
  Phys. Rev.} {\bf D73} (2006) 015004,
  [\href{http://xxx.lanl.gov/abs/hep-ph/0506023}{ hep-ph/0506023}].

\bibitem{Schmaltz:2006qs}
M.~Schmaltz and R.~Sundrum, {\it Conformal sequestering simplified},
  \href{http://xxx.lanl.gov/abs/hep-th/0608051}{ hep-th/0608051}.

\bibitem{Kaplan:1999ac}
D.~E. Kaplan, G.~D. Kribs, and M.~Schmaltz, {\it Supersymmetry breaking through
  transparent extra dimensions},  {\em Phys. Rev.} {\bf D62} (2000) 035010,
  [\href{http://xxx.lanl.gov/abs/hep-ph/9911293}{ hep-ph/9911293}].

\bibitem{Chacko:1999mi}
Z.~Chacko, M.~A. Luty, A.~E. Nelson, and E.~Ponton, {\it Gaugino mediated
  supersymmetry breaking},  {\em JHEP} {\bf 01} (2000) 003,
  [\href{http://xxx.lanl.gov/abs/hep-ph/9911323}{ hep-ph/9911323}].

\bibitem{Cheng:2001an}
H.~C. Cheng, D.~E. Kaplan, M.~Schmaltz, and W.~Skiba, {\it Deconstructing
  gaugino mediation},  {\em Phys. Lett.} {\bf B515} (2001) 395--399,
  [\href{http://xxx.lanl.gov/abs/hep-ph/0106098}{ hep-ph/0106098}].

\bibitem{Wells:2003tf}
J.~D. Wells, {\it Implications of supersymmetry breaking with a little
  hierarchy between gauginos and scalars},
  \href{http://xxx.lanl.gov/abs/hep-ph/0306127}{ hep-ph/0306127}.

\bibitem{Arkani-Hamed:2004fb}
N.~Arkani-Hamed and S.~Dimopoulos, {\it Supersymmetric unification without low
  energy supersymmetry and signatures for fine-tuning at the lhc},  {\em JHEP}
  {\bf 06} (2005) 073, [\href{http://xxx.lanl.gov/abs/hep-th/0405159}{
  hep-th/0405159}].

\bibitem{Drees:1986vd}
M.~Drees, {\it Intermediate scale symmetry breaking and the spectrum of super
  partners in superstring inspired supergravity models},  {\em Phys. Lett.}
  {\bf B181} (1986) 279.

\bibitem{Kolda:1995iw}
C.~F. Kolda and S.~P. Martin, {\it Low-energy supersymmetry with d term
  contributions to scalar masses},  {\em Phys. Rev.} {\bf D53} (1996)
  3871--3883, [\href{http://xxx.lanl.gov/abs/hep-ph/9503445}{
  hep-ph/9503445}].

\bibitem{Baer:1999mc}
H.~Baer, M.~A. Diaz, J.~Ferrandis, and X.~Tata, {\it Sparticle mass spectra
  from so(10) grand unified models with yukawa coupling unification},  {\em
  Phys. Rev.} {\bf D61} (2000) 111701,
  [\href{http://xxx.lanl.gov/abs/hep-ph/9907211}{ hep-ph/9907211}].

\bibitem{crs2}
A.~G. Cohen, T.~Roy, and M.~Schmaltz, {\it Scalar mass relations in the mssm
  with dynamical hidden sectors}, . to appear.

\bibitem{Giudice:1988yz}
G.~F. Giudice and A.~Masiero, {\it A natural solution to the mu problem in
  supergravity theories},  {\em Phys. Lett.} {\bf B206} (1988) 480--484.

\end{thebibliography}\endgroup
\bibliographystyle{JHEP}

\end{document}